\documentclass[prb,twocolumn,showpacs,preprintnumbers,amsmath,amssymb,showkeys,citeautoscript,floatfix]{revtex4}

\def \LOFSF{LaFeAsO$_{0.9}$F$_{0.1}$}
\def \Tc{$T_{\rm c}$}
\def \Tn{$T_{\rm N}$}
\def \Ts{$T_{\rm s}$}
\def \Bc2{$B_{\rm c2}$}

\def \ef{$f_{\rm 0}$}

\usepackage{graphicx}
\usepackage{bm}
\usepackage[dvips]{color}

\begin{document}
%
%
\title{Upper critical field, penetration depth and depinning frequency of the high-temperature superconductor~\LOFSF~studied by microwave surface impedance.}

%
\author{A. Narduzzo,$^{1}$ M. S. Grbi\'c,$^{2}$ M. Po\v{z}ek,$^2$ A. Dul\v{c}i\'{c},$^2$ D. Paar,$^{1,2}$ A. Kondrat,$^{1}$ C. Hess,$^{1}$\\ I. Hellmann,$^{1}$ R. Klingeler,$^1$ J. Werner,$^{1}$ A. K\"{o}hler,$^{1}$ G. Behr$^{1}$ and B. B\"{u}chner.$^1$}
\affiliation{$^1$Institute for Solid State Research, IFW Dresden, D-01101 Dresden, Germany;}
\affiliation{$^2$Department of Physics, Faculty of Science, University of Zagreb, P. O. Box 331, HR-10002 Zagreb, Croatia.}
\date{\today}
\begin{abstract}
Temperature and magnetic field dependent measurements of the microwave surface impedance of superconducting~\LOFSF~(\Tc~$\approx$~26K) reveal a very large upper critical field ($B_{\rm c2} \approx 56$T) and a large value of the depinning frequency ($f_{0}\approx 6$GHz); together with an upper limit for the effective London penetration depth, $\lambda_{\rm eff} \le 200 \rm \, nm$, our results indicate a
strong similarity between this
system and the high-$T_{\rm c}$ superconducting cuprates.
\end{abstract}
\maketitle
The recent discovery of superconductivity in \LOFSF~\cite{Kamihara:08} has lead to the rapid growth in the number of superconducting layered oxypnictides with larger and larger \Tc~($\approx$~55K in SmO$_{0.8}$F$_{0.2}$FeAs).~\cite{RenZA08LiuRHcm0804} Apart from their high \Tc, the interest in these materials stems primarily from the proximity of superconductivity to a spin-density-wave (SDW) ground state, and from the fact that multiple bands resulting from the orbitals of the conventionally pair-breaking magnetic ion Fe$^{2+}$ appear to be here directly responsible for the formation of the superconducting condensate.
Both {\it ab initio} band structure and LDA calculations\cite{Mazincm0803,Singhcm0803,Raghucm0804,Eremincm0804} show that the Fe-pnictide layers are responsible for the (super)conductivity; specifically, the $3d$ orbitals of the Fe$^{2+}$ ions weakly hybridized with the As$^{3-}$ $4p$ orbitals, form two electron and two hole pockets, while the [$RE$(OF)] ($RE$~=~La,Pr,Ce,Sm) layers act as charge reservoirs, F$^{1-}$ causing the electron doping, the size of the $RE$ element generating chemical pressure. The electronic structure thus consists of multiple quasi-two-dimensional Fermi surface sheets in the presence of competing ferromagnetic and antiferromagnetic fluctuations\cite{Singhcm0803,Eremincm0804}.
The microscopic nature of the superconducting pairing and the symmetry of the order parameter on the other hand are still far from established, with theoretical proposals ranging from extended s-wave mediated by antiferromagnetic spin fluctuations~\cite{Mazincm0803} to spin-triplet $p$-wave~\cite{Leecm0804}. The possibility of anomalously strong electron phonon coupling effects has also been emphasized~\cite{Eschrigcm0804}, while a very small value of the electron-phonon coupling constant ($\lambda_{\rm e-ph} < 0.21$ has been calculated~\cite{Boericm0803}.
The expected large moment for the undoped compound ($S=2$) is experimentally not observed, low temperature values of $\mu \approx 0.36\mu_{\rm B}$\cite{Delacruzcm0804} and $\mu \approx 0.25\mu_{\rm B}$\cite{Klausscm0805} being reported instead. In this parent compound a structural phase transition from tetragonal to orthorhombic occurs at \Ts~$\approx 150$K, closely related to the formation of a spin-density-wave (SDW) below \Tn~$\approx 135$K.
Electron doping rapidly suppresses both structural and SDW transitions leading to superconductivity, possibly allowing short-range magnetic fluctuations to survive in the region of the phase diagram where superconductivity becomes the preferred ground state. Whether or not these local fluctuations are responsible for the pairing is here the fundamental question. Further evidence for possible unconventional pairing with nodal order parameter comes from specific heat\cite{MuGcm0803}, tunneling\cite{ShanLcm0803}, magnetisation\cite{RenCcm0804}
and NMR\cite{Grafecm0805} measurements.

Here we report on temperature (4.2K~$<T$) and magnetic field dependence ($\mu_{\rm 0}H<16$T) of the microwave surface impedance of \LOFSF. The results allow us to estimate the upper critical field \Bc2, the effective London penetration depth $\lambda_{\rm eff}$ and the depinning frequency \ef~for this material.

Polycrystalline samples of \LOFSF~were prepared by a solid state reaction method and annealed in vacuum \cite{Drechslercm0805}. Inspection with a polarized light microscope revealed dense crystallites of sizes varying between 1 and 100 $\mu$m. The resistivity of the sample under study was measured by means of a standard DC method with four-point contact geometry and current polarity inversion. The magnetic susceptibility, both zero-field (ZFC) and field cooled (FC), was measured using a SQUID magnetometer.
The microwave measurements were carried out in a high-Q elliptical copper cavity at four different frequencies corresponding to four different resonant modes: the $_e$TE$_{111}$
mode at 9.1GHz, the $_e$TE$_{112}$ mode at 12.8GHz, the $_e$TE$_{211}$ mode at 15.1GHz and the $_e$TE$_{113}$ mode at 16.7GHz.
The sample was mounted on a sapphire sample holder and placed in the
center of the resonator.
In that position, the sample lies in a microwave {\it electric} field $E_{mw}$ maximum in modes $_e$TE$_{111}$ and $_e$TE$_{113}$ and in a microwave {\it magnetic} field $H_{mw}$ maximum in modes $_e$TE$_{112}$ and $_e$TE$_{211}$.
The temperature was varied between 5 and 50 K, and the applied DC magnetic field between 0 and 16 T.
Directly measured quantities are the $Q$-factor and the resonant
frequency $f$ of the cavity loaded with the sample.
The $Q$-factor was measured by a modulation technique described
elsewhere.\cite{Nebendahl:01} The empty cavity absorption
$(1/2Q)$ was subtracted as background from the measured data: the presented
experimental curves therefore display changes occurring exclusively in the physical properties of the samples themselves. An automatic frequency control (AFC) system was used to track the
source frequency always in resonance with the cavity. Thus, the
frequency shift could be measured as the temperature of the sample or the static magnetic field were
varied.
The two measured quantities represent the complex frequency shift
$\Delta \widetilde{\omega} / \omega = \Delta f/f  + i \Delta (1/2Q)$.
\begin{figure}[h]
\includegraphics[width=6.7cm]{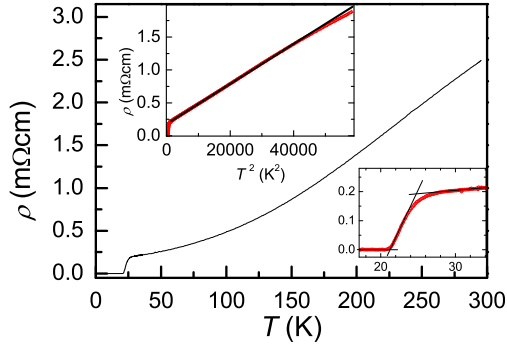}%
\caption{(Color online) temperature dependence of the resistivity; top left hand side inset: resistivity plotted versus $T^2$, showing a deviation above $T \approx 200$K; bottom right hand side inset: the superconducting transition.}
\label{rho}
\end{figure}
\begin{figure}[h]
\includegraphics[width=6.4cm]{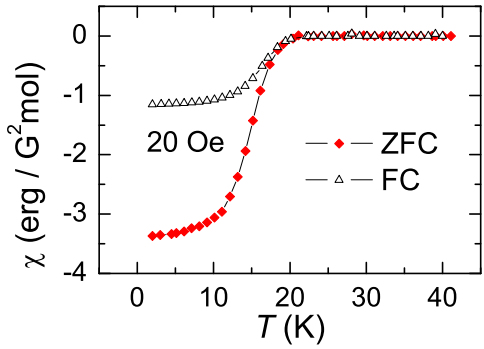}%
\caption{(Color online) temperature dependence of the susceptibility measured in zero field (ZFC) and in applied magnetic field (FC, 20Oe).}
\label{chi}
\end{figure}

\begin{figure}[h]
\includegraphics[width=7.6cm]{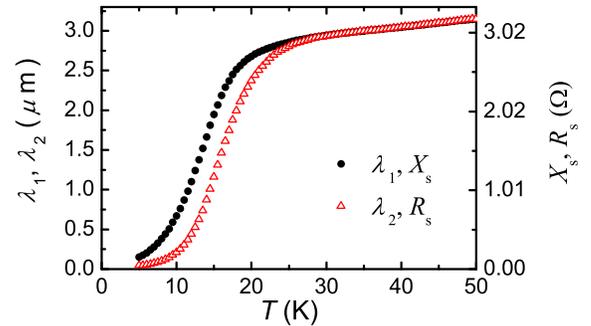}%
\caption{(Color online) temperature dependence of the complex penetration depth (left axis) and the complex impedance (right axis) in zero applied magnetic field.}
\label{Lambda}
\end{figure}

\begin{figure}[h]
\includegraphics[width=6.2cm]{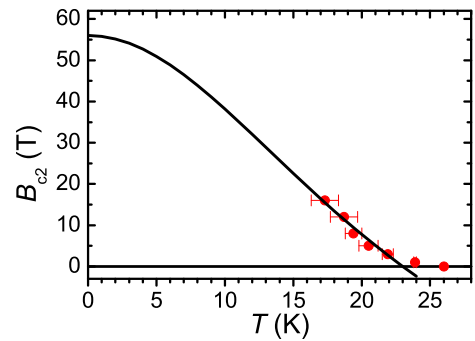}%
\caption{(Color online) the upper critical field as extracted from the temperature dependence of the complex conductivity in an applied magnetic field; the line is a fit to the data below \Tc=23.7K using equation~\ref{Bc2_GL}.}
\label{Bc2}
\end{figure}

The temperature dependence of resistivity and susceptibility are shown in Fig.~\ref{rho} and
Fig.~\ref{chi}~respectively.
Remarkably, the normal state resistivity has a quadratic temperature dependence up to about 200K. The midpoint of the resistive transition yields \Tc=23.7K with a width $\Delta T \approx 4$K  (90\% - 10\% criterion),
while the onset of diamagnetism from the susceptibility curve (ZFC) becomes discernible below $T \approx 22$K.
Both data sets reveal a sample with some degree of inhomogeneity, with a fluorine content slightly different from the nominal value of 0.1.
The field-cooled (FC) susceptibility shows significant flux penetration for fields as low as 20 Oe.
\begin{figure}[h]
\includegraphics[width=6.7cm]{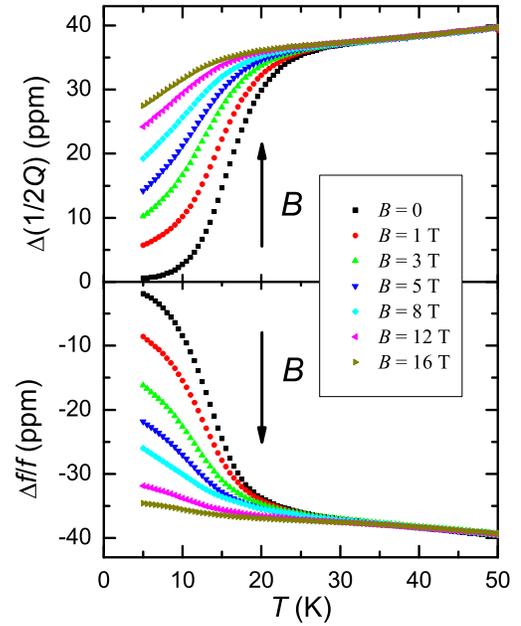}%
\caption{(Color online) temperature dependence of the complex frequency shift in various applied magnetic fields: imaginary part in the top panel and real part in the bottom panel.}
\label{ComplexShift}
\end{figure}
The two panels of Fig.~\ref{ComplexShift} show the measured complex frequency shift for various values of the applied DC magnetic field in the microwave mode $_e$TE$_{112}$. For a thick sample there is a proportionality between the complex frequency shift and the surface impedance:
$Z_s \propto -i \Delta \widetilde{\omega} / \omega$.
The factor of proportionality can be determined from the normal state resistivity $\rho_n (T=30 \rm K) =(0.20\pm 0.05)\, \rm m\Omega cm$.
From the surface impedance one can determine the complex penetration depth $\widetilde{\lambda}=\lambda_1 - i \lambda_2$ through\cite{Coffey:91}
\begin{equation}
Z_s=i \mu_0 \omega \widetilde{\lambda} \, \,.
\end{equation}
The resulting temperature dependencies of $\lambda_1$ and $ \lambda_2$ in zero applied magnetic field are shown in Fig.~\ref{Lambda}.
In the normal state one has $\lambda_1 = \lambda_2 = \delta_n /2$, where $\delta_n$ is the normal metal skin depth.
In the opposite limit ($T \rightarrow 0$), where the real part of the complex conductivity $\widetilde{\sigma}=\sigma_1 - i \sigma_2$
disappears, the complex penetration depth becomes real and identical to the London penetration depth $\lambda_L = \lambda_1(T=0)$.
This analysis was performed for all four microwave modes leading to an estimate of the penetration depth at $T=5\, \rm K$ to be $\lambda_1=(200\pm 50) \rm nm$.
This is an effective value for the polycrystalline sample and can be taken
as the upper limit of the intrinsic value of the London penetration depth
in the $ab$ plane. It is clear from Fig.~\ref{Lambda} that $\lambda_1 (T)$
does not saturate at 5 K; for a nodal order parameter, it would
have a linear dependence at very low temperatures\cite{Hardy:93} due to
the gradual excitation of quasiparticles from the superconducting
condensate. By linearly extrapolating $\lambda_1(T)$ down to 0K,
a value substantially smaller than 200 nm for the zero temperature London
penetration depth would be obtained.
From the data in Fig.~\ref{ComplexShift} the upper critical field \Bc2 can be estimated.
An empirical criterion for the onset of superconductivity at a given applied field would be the deviation from the apparently linear (normal state) behaviour of the absorption in the top panel of Fig.~\ref{ComplexShift}. This method, however, is not very precise, and we therefore decided to apply a quantitative, arguably more rigorous, way of determining such a point. From the complex frequency shift one can extract the complex conductivity $\widetilde{\sigma}=\sigma_1 - i \sigma_2$; the emergence of $\sigma_2$ is the sign of the establishment of the coherent superconducting state.
In Fig.~\ref{Bc2} we have plotted the upper critical fields determined by the criterion that $\sigma_2$ exceeds 1\% of the normal state $\sigma_n$. The solid line is the plot of the following formula derived from Ginzburg-Landau theory\cite{Tinkham}
\begin{equation}
B_{\rm c2}(T)=B_{\rm c2}(0) \frac{1-t^2}{1+t^2} \, \, ,
\label{Bc2_GL}
\end{equation}
where $t=T/T_0$, with $T_0=23.0\, \rm K$ and $B_{\rm c2}(0)=56\, \rm T$. The slope of \Bc2$(T)$ near \Tc, $dB_{\rm c2}(T)/dT$, is $-2.5$T/K; this value is substantially unaffected if a different criterion (percentage) for the onset of superconductivity from microwave measurements is adopted.
We have neglected in the fit the points above \Tc$=23.7$K, the critical temperature as deduced from resistivity measurements. Taking these points into account (points that may possibly only be representative of crystallites with a slightly higher~\Tc) would lead to a somewhat smaller value of \Bc2 but more importantly to an underestimate of the slope of \Bc2$(T)$ near~\Tc.
We have also measured the field dependence of the complex frequency shift at several constant temperatures.
We could not observe the low field signal changes typical of intergrain
Josephson junctions. It therefore appears that the preparation of the sample under 1GPa pressure resulted in a very compact granular structure.
From this frequency shift the mixed state effective complex conductivity can be analytically extracted.
The effective conductivity\cite{Coffey:91,Brandt:91,Dulcic:93} in an oscillating electric field is given by:
\begin{equation}
\frac{1}{\widetilde{\sigma}_{\mathrm{eff}}}= \frac{1-\frac{b}{1-i(\omega_0/\omega)}}
{(1-b)(\sigma_1-i \sigma_2)+b \sigma_n}+
\frac{1}{\sigma_n}\, \frac{b}{1-i(\omega_0/\omega)} \, \, .
\label{sigmaeff}
\end{equation}
The parameter $b$ represents the volume fraction of the sample occupied by the normal vortex cores. $\omega_0$ is the depinning frequency which depends on sample, field and temperature, ranging from the
strongly pinned case ($\omega_0  \gg \omega$) to the flux flow limit ($\omega_0 \ll \omega$), where $\omega$ is the driving microwave frequency.
\begin{figure}[h]
\includegraphics[width=6.5cm]{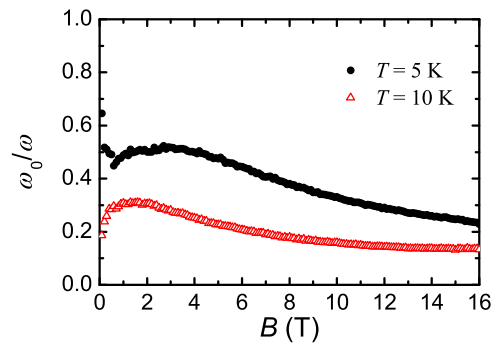}%
\caption{(Color online) the field dependence of the depinning frequency $\omega_0$ at two temperatures.
The driving frequency was $\omega = 2\, \pi\, 12.77 \rm GHz$.}
\label{Omega}
\end{figure}
By numerical inversion of Eq.~(\ref{sigmaeff}), we have determined the
values of $b$ and $\omega_0$. For the driving frequency $\omega = 2\, \pi\, 12.77 \rm GHz$, the field dependence of $\omega_0 / \omega$ is plotted in Fig.~\ref{Omega}.
The highest value of the depinning frequency $f_0=\omega_0 /(2 \pi)$ is obtained for $T=5\, \rm K$ at
low fields: $f_0 \approx 6\, \rm GHz$.
With increasing field, and/or temperature, $f_0$ decreases. Thus, most of the microwave measurements are close to
the flux-flow regime.
%

The normal state resistivity reveals a reasonably good metal ($\rho_0 \approx 200 \mu \Omega$cm, RRR=$\rho$(300K)/$\rho$(30K)=12.1) with $\rho \sim T^2$ up to $T \approx 200$K. This feature is remarkable, but indeed more experimental evidence is needed to prove this to be a signature of Fermi liquid behavior.
Our measured effective penetration depth (an upper limit to the London penetration depth) is somewhat smaller than the value of $\lambda_{\rm ab}(0)$ measured by $\mu$SR experiments on \LOFSF\cite{Luetkenscm0804};
with a value for $\lambda_{\rm eff}^{-2}\ge 25 \mu \rm m^{-2}$, this would position our compound closer than previous results to the line of the electron-doped cuprates on the Uemura plot\cite{Luetkenscm0804,Drewcm0805}.
Note that our measurement does not rely on any assumption regarding the distribution and arrangement of the vortex lattice within the sample in order to extract $\lambda_{\rm eff}$.
The obtained values of upper critical field, \Bc2=56T, and slope near \Tc, $dB_{\rm c2}(T)/dT=-2.5$T/K, are in substantial agreement with other resistivity measurements: for compounds with nominally
the same doping, Zhu {\it et al} obtain the same value of \Bc2 and a similar slope ($dB_{\rm c2}(T)/dT=-2.3$T/K) using formula~\ref{Bc2_GL} for their fit\cite{Zhu0803}, while Hunte {\it et al} obtain \Bc2 in the range 62-65T and a similar slope by applying the conventional one-band Werthamer-Helfand-Hohenberg theory\cite{Huntecm0804}. Their result is closer to that reported by Fuchs {\it et al} on As-deficient \LOFSF~samples\cite{Fuchscm0806}, whose value of $dB_{\rm c2}(T)/dT$ near \Tc~though is considerably larger. The measurement of the depinning frequency yields
$f_0\approx 6$GHz: a number well into the microwave range.
Typically, copper based high-$T_{\rm c}$ superconductors have depinning frequencies slightly higher than 10GHz\cite{Golosovsky:94}, while the depinning frequencies in classical bulk superconductors are below 100MHz\cite{Gittleman:66}. This result therefore also points to a substantial communality of features
between these novel materials and the cuprates.

In summary, microwave surface impedance measurements on the novel superconductor~\LOFSF~ provide estimates of
the upper critical field (\Bc2=56T) and the penetration depth ($\lambda_{\rm eff}\le 200$nm);
the latter appears to be substantially smaller than the values estimated from measurements carried out by other techniques.\cite{Zhu0803,Luetkenscm0804}.
Together with the large value of the depinning frequency ($f_0\approx 6$GHz), these results yield a phenomenological picture of this system that closely resembles that of the high-$T_{\rm c}$ cuprate superconductors.

We thank S.-L. Drechsler, G. Fuchs and I. Vekhter for their valuable comments.
We acknowledge financial support from the Croatian Ministry of Science,
Education and Sports (project no. 119-1191458-1022 ``Microwave Investigations
of New Materials'').

\end{document}